\documentclass[sigconf]{acmart}

\usepackage{booktabs} 
\usepackage{graphicx}
\usepackage{subcaption}
\usepackage{caption}
\usepackage{tabularx}
\usepackage{xspace}
\usepackage{bera}
\usepackage{listings}
\usepackage{xcolor}
\usepackage{enumitem}
\usepackage{breakurl}
\usepackage{pifont}

\colorlet{punct}{red!60!black}
\definecolor{background}{HTML}{EEEEEE}
\definecolor{delim}{RGB}{20,105,176}
\colorlet{numb}{magenta!60!black}

\lstdefinelanguage{json}{
    basicstyle=\normalfont\ttfamily,
    numbers=left,
    numberstyle=\scriptsize,
    stepnumber=1,
    numbersep=8pt,
    showstringspaces=false,
    breaklines=true,
    frame=lines,
    backgroundcolor=\color{background},
    literate=
     *{0}{{{\color{numb}0}}}{1}
      {1}{{{\color{numb}1}}}{1}
      {2}{{{\color{numb}2}}}{1}
      {3}{{{\color{numb}3}}}{1}
      {4}{{{\color{numb}4}}}{1}
      {5}{{{\color{numb}5}}}{1}
      {6}{{{\color{numb}6}}}{1}
      {7}{{{\color{numb}7}}}{1}
      {8}{{{\color{numb}8}}}{1}
      {9}{{{\color{numb}9}}}{1}
      {:}{{{\color{punct}{:}}}}{1}
      {,}{{{\color{punct}{,}}}}{1}
      {\{}{{{\color{delim}{\{}}}}{1}
      {\}}{{{\color{delim}{\}}}}}{1}
      {[}{{{\color{delim}{[}}}}{1}
      {]}{{{\color{delim}{]}}}}{1},
}

\setcopyright{none}
 \renewcommand\footnotetextcopyrightpermission[1]{} 
\newcommand{\name}{\textsf{IDIoT}\xspace}

\begin{document}
\title{IDIoT: Securing the Internet of Things like it's 1994}

\author{David Barrera}
\affiliation{
  \institution{Polytechnique Montr\'{e}al}
}

\author{Ian Molloy}
\author{Heqing Huang}
\affiliation{
  \institution{IBM T.J. Watson Research Center}
}

\begin{abstract}

Over 20 billion Internet of Things devices are set to come online by 2020. Protecting such a large number of underpowered, UI-less, network-connected devices will require a new security paradigm. We argue that solutions dependent on vendor cooperation such as secure coding and platform changes are unlikely to provide adequate defenses for the majority of devices. Similarly, regulation approaches face a number implementation challenges which limit their effectiveness. As part of the new paradigm, we propose \name, a network security policy enforcement framework for IoT devices. \name prevents widespread network attacks by restricting IoT devices to only their necessary network behavior. \name is simple and effective, building on decades of tried-and-true network security principles without requiring changes to the devices or cloud infrastructure.

\end{abstract}

\maketitle

\section{Introduction}

In the fall of 2016, several high-profile websites (Github, Twitter, Reddit, Netflix, and others) were taken offline by a series of distributed denial of service (DDoS) attacks. The DDoS attacks post-mortem revealed that the botnet code (later dubbed \textit{Mirai}) was designed to discover and compromise embedded devices using default credentials. Further analysis estimated that 150,000 embedded devices (digital video recorders and surveillance cameras) were involved in the attack~\cite{securityweek}. 

The Mirai attacks were noteworthy for two reasons: (1) the attack volume (up to 1.5 Tbps) was one of the largest to date, and was powerful enough to take down both a large cloud service provider~\cite{ovh} and a major DNS provider~\cite{dyn}, resulting in Internet-wide outages. (2) The botnet provided clear evidence that Internet of things (IoT) devices could be leveraged as botnet clients, despite these embedded devices not serving as general purpose computers. 

Following Mirai, more IoT attacks were discovered. The Vault 7 leaks~\cite{vault7} showed that state-level adversaries had developed malware for smart TVs, which allowed them to use the TV as covert spying devices. In April, the \emph{Brickerbot} malware was reportedly infecting IoT devices and permanently disabling them. The malware infected devices via \verb|telnet|, and overwrote the content of storage devices with random bytes~\cite{brickerbot}.

The rise in IoT-specific attacks is perhaps unsurprising. As the cost of adding wireless capabilities to devices decreases, more consumer electronics, toys, appliances, and other ``things'' are becoming Internet-enabled. A 2015 report~\cite{gartneriot} estimates that by 2020, there will be around 20 billion IoT devices online, with 65\% of those existing in the consumer space. IoT devices exhibit desirable characteristics for attackers: they are powered-on and networked 24/7, they use weak security configurations, and they are underpowered so they cannot run anti-malware, intrusion detection, or auditing services. Software or firmware update procedures for IoT devices tends to be clunky and error-prone, allowing devices to remain unpatched for long periods of time. Moreover, embedded devices often lack displays, making it difficult to know whether the device is behaving as expected at any given time.

This paper explores the new paradigm of securing 20 billion+ Internet of Things devices. We discuss challenges and assumptions that complicate the deployment and use of proposed solutions. We suggest, counter-intuitively, that solving the IoT security problem may not necessarily require complex new hardware and software stacks. Instead, returning to tried-and-true principles of network security can be effective in securing devices. Our insight is that unlike personal computers, IoT devices tend to behave predictably, and are thus well-suited for network isolation and filtering techniques. Correct use of these techniques allows even vulnerable devices to continue regular operation, co-existing with other devices on the network. 

We demonstrate these ideas through our implementation of \name, a network security policy enforcement architecture for IoT devices. \name filters the outgoing network connections of IoT devices, dropping those that are not required for essential device operation. For example, surveillance cameras should not be allowed to send traffic to Twitter or Reddit, but should be allowed to upload video streams to a cloud storage provider. These rules are encoded in a policy, which is enforced close to the IoT devices. 

\smallskip
\noindent\textbf{Scope.} Some definitions of IoT include industrial control systems, smart-grid devices like smart meters, and vehicular systems. In this paper, we focus on \emph{consumer} IoT -- devices that an end user could purchase and connect to their home network.

\section{Background and Related Work}

\subsection{Surveys}
Despite being relatively new, the IoT security research space already has several surveys describing protocols, architectures, design patterns, and research gaps. A 2010 non-security focused survey~\cite{atzori_internet_2010} envisioned the IoT as a set of devices that can communicate directly (without requiring the Internet, but still requiring a local area network). Under this many-to-many communication model, the survey highlighted challenges in network addressing, protocol standardization, and security. However, the security issues described do not appear to be IoT specific, focusing instead on high-level issues such as authentication, data integrity, and privacy.  

In 2013, Roman et al.~\cite{roman_features_2013} compared the distributed IoT paradigm to
a centralized approach describing the different security and privacy
requirements of each model. The paper noted that while a distributed IoT model
encourages privacy (data doesn't have to be surrendered to a cloud provider to
obtain the desired functionality), deployment challenges such as fault tolerance
and interoperability remain unsolved. Interoperability is discussed in detail by
Keoh et al.~\cite{keoh_securing_2014} and Granjal et
al.~\cite{granjal_security_2015}, who surveyed the myriad of IETF-proposed
standards describing protocols for IoT communication at the physical (PHY),
media (MAC), routing, and application layers. Standardization efforts appear to
be focusing on DTLS (TLS over UDP) as the secure end-to-end transport layer, but
there are still open questions around strategies for provisioning or rotating key
material, how to deprecate old or add support for new cryptographic primitives,
and what key lengths are optimal given some devices' limited processing power.

\subsection{Security Analysis of IoT Devices}
This section reviews academic work focusing on the security analysis of embedded and IoT devices. We discuss large-scale evaluations of multiple device types, followed by research that focuses on specific devices such as Internet-enabled locks, networked printers, and smart lighting. 

\smallskip
\noindent\textbf{Large-scale analysis.} A 2010 study~\cite{cui_quantitative_2010} scanned the IPv4 address space and found that 13\% of devices (about 540,000) that responded were vulnerable to compromise due to the use of default credentials. While the highest ratio of vulnerable devices was seen on ISP-issued modems/routers, the scan also turned up IoT devices: DVRs, cameras, and VoIP appliances. Cameras with default credentials were the culprit in the recent Mirai botnet attack~\cite{mirai} which took down major DNS resolvers. Pa et al.~\cite{pa_iotpot:_2015} developed a honeypot to monitor botnets that target IoT devices. The analysis revealed four botnet families targeting devices via \verb|telnet|, and after compromise using those devices to launch further attacks.  Default credentials and \verb|telnet| remain active forms of exploitation of IoT devices; in March 2017, security firm Radware discovered the \emph{Brickerbot} worm. Brickerbot spreads via \verb|telnet| and corrupts storage volumes on victim devices~\cite{brickerbot}.

Costin et al.~\cite{andrei_costin_large-scale_2014} analyzed 32,000 embedded device firmware images for printers, routers, cameras, etc. Without prior knowledge of the firmware image layout or access to the device for which the firmware was developed, the authors were able to extract 35,000 RSA private keys, weak password hashes, and hardcoded credentials. The authors also found a number of vendor-installed backdoors via the SSH \verb|authorized_keys| directive.

Heninger et al.~\cite{heninger_mining_2012} analyzed TLS certificates and SSH host keys collected from Internet-wide scans. They found vulnerabilities resulting from the use of default or well-known private keys, duplicate keys due to insufficient entropy during key generation, and were able to factor private keys due to insufficient signature randomness or shared common factors. While most of the vulnerable devices were enterprise-grade network devices, they also found consumer routers and VoIP products, and network-attached storage devices had similar vulnerabilities. Heninger et al. demonstrated that even the correct use of authentication systems can become vulnerable if the underlying operating system or hardware fails to provide sufficient entropy. 

\smallskip
\noindent\textbf{Smart Locks.} The security of several IoT smart locks was evaluated by Ho et al.~\cite{ho_smart_2016}. The paper outlines the challenges in securing the smart lock architecture, which may involve polling a cloud service to retrieve an updated list of access control rules. While the findings and suggestions in their work are specific to smart locks, some of the findings are more broadly applicable to other IoT devices. For example, the authors suggest eliminating reliance on cloud services for critical functionality, since cloud interactions may not always be available (e.g., server crash or the company goes out of business) and when available add latency which decreases usability. 

\smallskip
\noindent\textbf{Printers.} A survey on the security of network-connected printers revealed the presence of decades-old vulnerabilities in recent devices, as well as new vulnerabilities that become possible by Internet-enabled printing~\cite{jens_muller_sok:_2017}. The survey documents the challenges in securing devices that must accept arbitrary code for their basic functionality, as well as manufacturers' unwillingness/inability to provide basic levels of security.

\smallskip
\noindent\textbf{Smart Light Bulbs.} By exploiting a bug in the low-power local area network protocol used by Philips smart lightbulbs (Zigbee), Ronen et al.~\cite{ronen_iot_2017} were able to control victim light bulbs from a distance. Moreover, Ronen et al.~used a side-channel attack to recover the AES key used by Philips to authenticate and encrypt firmware updates, allowing them to create and deploy malicious firmwares. 

\subsection{Network Isolation and Filtering}
\label{sec:enterprisenetsec}

To prevent access of rogue devices onto their networks, enterprise network administrators have traditionally deployed network access control (NAC) solutions. NAC allows access to a wired or wireless network only if a client passes a set of security tests pre-defined in a policy. The security tests can vary from simply proving ownership of valid credentials (e.g., for accessing premium hotel Wi-Fi), to comprehensive system inspections verifying the freshness of AV signatures, the presence or absence of software packages, the versions of specific packages, among others. 

Comprehensive NAC solutions, where administrators provision the devices that are given to employees (e.g., laptops, mobile phones), use client-side \textit{agents}, which report system status and integrity measurements. When installing an agent is not possible, the NAC may use captive portals to allow web-based authentication. Academic NAC proposals have suggested using behavioral profiles instead of manually defined policies to allow access to clients~\cite{frias2008,frias2009}. However, these dynamic approaches may still incorrectly allow access to the network to devices that can fool classifier. 

It appears existing NAC solutions aren't well suited for IoT devices since these devices cannot run agents, or authenticate via captive portals. Additionally, the all-or-nothing access to the network approach is too coarse grained, failing to offer protection for devices that were previously authorized to join the network, but later become compromised. 

In recent years, a variety of integrated network security solutions have emerged with the focus of protecting corporate networks. Palo Alto Networks, Fortinet, Barracuda Networks, and other companies now sell products marketed as \textit{next-generation firewalls (NGFW)} and \textit{unified threat management (UTM)} appliances. In addition to basic packet filtering, these security products integrate network monitoring, user/application awareness, intrusion prevention, anti-virus, SSL interception, captive portals, spam detection, etc. These features, along with attractive UIs and remote management capabilities, come at high premium; a next-generation firewall can cost upwards of US\$1000 and require per-user licensing fees and support contracts. These high costs are likely beyond acceptable for most households, and the simplicity of many IoT devices makes NGFW and UTM features overkill. 

\section{Proposals for Securing the Internet of Things}

This section broadly collects ideas and suggestions for securing the Internet of Things. We do not aim to provide a comprehensive list of all proposals to date, but rather we aim to categorize, exemplify, and challenge representative ideas found in the academic literature and news articles. 

\subsection{General Guidelines}

\subsubsection{Use of standard protocols.} The idea here is that if vendors make use of standard or well-understood protocols, rather than custom solutions, they are less likely to create insecure devices. Unfortunately the community has yet to reach agreement on which of the dozens of standards to use for each layer of the IoT stack~\cite{granjal_security_2015}. Even if vendors agree on a set of standards, these may still allow insecure operation. For example the Constrained Application Protocol (CoAP) -- specifically designed for IoT devices -- allows a \textit{NoSec} parameter, which when used transmits data without using authentication, encryption, or integrity protection. When \textit{NoSec} is used, these data protection mechanisms should be implemented at the application layer.

\subsubsection{Pen-testing/Code review for devices.} While such techniques are generally regarded as software development best practice~\cite{mcgraw2004software}, pen-testing tends to focus on the identification of known vulnerabilities. Additionally, once the \emph{honeymoon period}\footnote{Clark et al.~\cite{clark_familiarity_2010, clark_blood_2010} define the honeymoon period of software as the period after a release during which no vulnerability has been found in a given software product.} ends, the time between the discovery of subsequent vulnerabilities decreases. 

Code review may help detect issues prior to release, but this practice is not always effective, especially for small development teams. Academic efforts exist to develop scalable automated testing of IoT devices~\cite{rosenkranz_distributed_2015}, but the challenges introduced by hardware, software, and toolchain heterogeneity limit the chance of success of such one-size-fits-all testing frameworks.

\subsection{Technical Changes}

While the security literature offers a rich set of secure programming languages, architectural/platform security solutions, and even verifiable operating systems (e.g., seL4~\cite{klein2009sel4}), these tools remain costly to deploy. The economic incentives for IoT vendors to use and maintain such solutions, much like in the case of secure desktop software~\cite{anderson_why_2001}, don't appear to exist. Given the rapid pace of innovation in IoT, it seems to be more profitable to be first than to be secure.

In the hardware space, new architectures (e.g., Sancus~\cite{job_noorman_sancus:_2013}) aim to provide a hardware-only trusted computing base (TCB) for developing applications on embedded systems. While it is clear that such an architectures can offer stronger security properties for embedded systems than off-the-shelf platforms, it is less clear that such systems would be used correctly if pervasively deployed. For example, Sancus offers no mechanism for detecting or revoking compromised cryptographic keys, forcing developers to come up with their own solutions to these seemingly fundamental issues. 

To make matters worse, research on side-channel attacks~\cite{degabriele_provable_2011} has shown repeatedly that even when systems are protected through correct use of cryptographic systems, those protections can still be bypassed. While correct use of cryptographic systems may force adversaries to higher-cost attacks, it is becoming increasingly evident that no assumptions can be made about the long-term security of code, protocols, or devices. 

In the network space, Yu et al.~\cite{yu_handling_2015} proposed software-defined networking (SDN) architecture to secure inter-IoT device communication. The proposal, called \textit{IoTSec}, adds a virtual middlebox between each IoT device on the network and the gateway. At each middlebox, a high-level policy (i.e., defining allowed application-layer interactions rather than packets or protocols) is installed, which defines a set of allowed interactions between a protected device and other devices on the network. IoTSec requires that home networks be re-architected to support SDN, and that all possible cross-device interactions be enumerated in order to create the policies. 

\subsection{Regulation}

From a non-technical standpoint, there has been an increasing call to regulate the IoT space~\cite{waporegulations}, citing market failures as the primary reason for which IoT vendors will not independently secure their products. While regulation appears necessary, concrete examples of how regulation can help IoT security remain to be seen. Moreover, regulation has a number of challenges:

\begin{itemize}[leftmargin=*]

\item \textbf{What regulations are actionable by vendors?} Requiring more secure defaults is something that can be implemented easily by vendors. However, weak default credentials and configurations are not the only source of attacks. More complex vulnerabilities arising from buffer overflows or weak entropy are difficult to solve prior to release, as is known from the desktop software space. 

\item \textbf{Who becomes the regulator?} There are challenges in selecting or forming a third-party verifier who can give a security stamp of approval. This is particularly challenging and costly to enforce across borders. Additionally, lack of consistency across verifiers (e.g., different definitions of ``secure'') can lead to confusion for consumers~\cite{spectrumregulations}.

\item \textbf{How long are products regulated?} With IoT devices acting as simple switches, sensors, and toys, many of these will outlive warranty and support periods with some even outliving the company that produced them. It is unclear if security regulation can be effective or enforced beyond the support period, giving vendors incentive to obsolete their products more rapidly. 

\item \textbf{Compliance.} As has been seen with regulation in other domains (e.g,. PCI-DSS, HIPAA, automotive emissions), imposing a set of minimum security standards often leads to vendors complying with exactly those minimum standards, since every additional security feature or system has additional cost. That additional cost will typically not be recovered since security is not a differentiator in the consumer space~\cite{anderson_why_2001}.

\end{itemize}

\begin{table}[]
\centering
\begin{tabular}{@{}p{3.7cm}ccc@{}}
\toprule
& \textbf{Distinct} & \textbf{Distinct} & \textbf{HC} \\ \textbf{Device} & \textbf{Endpoints} & \textbf{Domains} & \textbf{IPs} \\ \midrule
AT\&T Microcell & 2 & 0 & 2    \\
Fitbit Aria Digital Scale & 2 & 1 & 0 \\
Withings Smart scale$\dagger$ & 2 & 1 & 0  \\
Withings Baby Monitor$\dagger$ & 2 & 1 & 0 \\
PIX-STAR Photo-frame$\dagger$ & 2 & 1 & 0 \\
Belkin Wemo switch$\dagger$ & 2 & 1 & 0  \\
Blipcare BP meter$\dagger$ & 2 & 1 & 0 \\
Samsung Bluray Player & 4 & 1 & 0 \\
Netatmo Weather Station & 5 & 1 & 0 \\
LIFX Gen 1 bulb\ding{83} & 5 & 1 & 0 \\
LIFX Gen 2 bulb\ding{83} & 5 & 2 & 0 \\
Triby Speaker$\dagger$ & 6 & 2 & 0 \\
NEST Smoke Alarm$\dagger$ & 6 & 4 & 0 \\
TP-Link Smart plug$\dagger$ & 7 & 2 & 0 \\
Netatmo Welcome$\dagger$ & 7 & 2 & 6 \\
Amazon Fire TV & 8 & 4 & 0 \\
Amazon Kindle & 9 & 8 & 1 \\
TP-Link Cloud camera$\dagger$ & 15 & 2 & 3 \\ 
Amazon Echo\ding{83} & 20 & 13 & 0 \\ \bottomrule
AppleTV 4th Gen & 37 & 23 & 2  \\
Samsung Galaxy Tab$\dagger$\ding{83} & 48 & 21 & 0 \\
Android Phone$\dagger$ & 57 & 48 & 0 \\ 
Microsoft XBox One & 74 & 57 & 0  \\ 
Laptop$\dagger$ & 140 & 101 & 0 \\ \bottomrule
\end{tabular}
\caption{Network behavior of several IoT devices. General purpose computing systems given in the bottom rows for comparison. HC IPs are hardcoded IP addresses, marked if no corresponding DNS lookup was made prior to connecting to an IP address. Devices with \ding{83} ignored the DHCP-provided DNS resolver and used Google's resolver (8.8.8.8) instead. Data for devices with $\dagger$ was obtained from the public dataset of Sivanathan et al.~\cite{dataset}.}
\label{tab:summary}
\end{table}

\section{The \name Policy Enforcement System}

Today's consumer IoT devices do not behave like general purpose computers. The lack of graphical or other interfaces on most IoT devices prevents users from directly running their own software on the devices, even though the underlying operating system may support it. 

Table~\ref{tab:summary} lists a summary of the network behavior of 19 IoT devices. To populate this table, we monitored network traffic of each device for approximately 12 minutes beginning with device power-on. We avoided interacting with device during the capture period, which gives us a baseline of network activity done automatically by devices at boot time. Our dataset was augmented with the public dataset of Sivanathan et al.~\cite{dataset}. Our analysis shows that simple devices such as digital scales, smart light bulbs, and Bluray players have a small network footprint. These devices look up a small set of domain names (typically API endpoints and domains used for network connectivity checks), and only connect to the servers returned by the DNS lookups of those domain names. By contrast, more complex devices allowing installation of apps (e.g., laptops, mobile phones, and game consoles) connect to a larger set of remote hosts and perform more DNS lookups.

Current home networks treat IoT devices as general purpose computing systems, allowing unrestricted network access to the devices despite only requiring a small set of connections to support their functionality. This over-privilege creates an opportunity for attackers to use victim IoT devices to launch other attacks (e.g., the Mirai DDoS attacks).

To mitigate this threat, we designed \name, a network policy enforcement architecture for consumer IoT devices. The goal of \name is to restrict the network capabilities of IoT devices to \emph{only what is essential for regular device operation}. For example, an Internet-connected surveillance camera should be allowed to upload video feeds to a cloud service, but should \emph{not} be permitted to send arbitrary UDP traffic to DNS resolvers. \name policies allow the specification of network activity that should be allowed, while any traffic not specified in the policy is dropped at a policy enforcement point.  

\subsection{Design Goals}
Our design of \name is informed by empirical observations about current IoT devices and threats to those devices. We additionally seek to achieve the following goals: 

\begin{itemize}[leftmargin=*]
\item \textbf{Deployability.} The system should provide security benefits without requiring changes to either the IoT devices or to the cloud services that that support them. Additionally, the system should support standalone operation without requiring vendor or third-party support (but could benefit from such support, as explained later).

\item \textbf{Extensibility.} With the growing applications of IoT, it is unreasonable to expect a solution designed for today's devices will work for all future devices as well. The system should support updates to policies and enforcement as devices and technologies evolve. Moreover, updates providing support for new devices, new filtering techniques, or new policies should not worsen existing protection to devices.

\item \textbf{Simplicity.} It is possible that heavyweight enterprise-grade appliances (see Section~\ref{sec:enterprisenetsec}) could offer similar levels of protection. However, we expect most consumers to lack interest, ability, and resources to deploy enterprise-grade hardware in their households. The system should build upon simple, yet effective, network filtering techniques, and avoid unnecessary use of anomaly detection or other techniques that could misclassify traffic.

\end{itemize}

\subsection{Overview}

\begin{figure}
\centering
\begin{subfigure}{.5\textwidth}
  \centering
  \includegraphics[width=.7\textwidth]{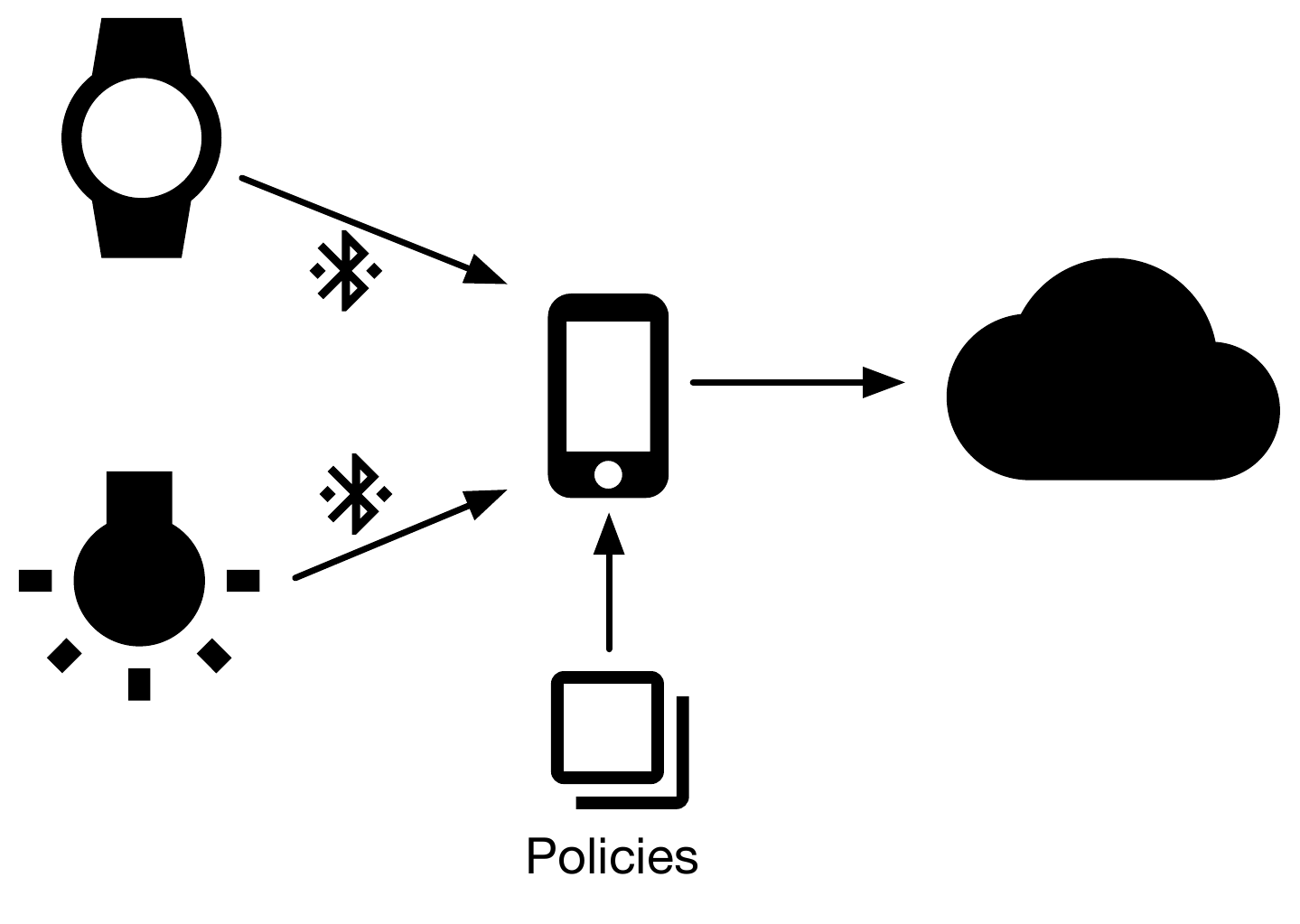}
  \caption{Enforcement at a smartphone}
  \label{fig:overview-phone}
\end{subfigure}\\%
\vspace{2em}
\begin{subfigure}{.5\textwidth}
  \centering
  \includegraphics[width=.8\textwidth]{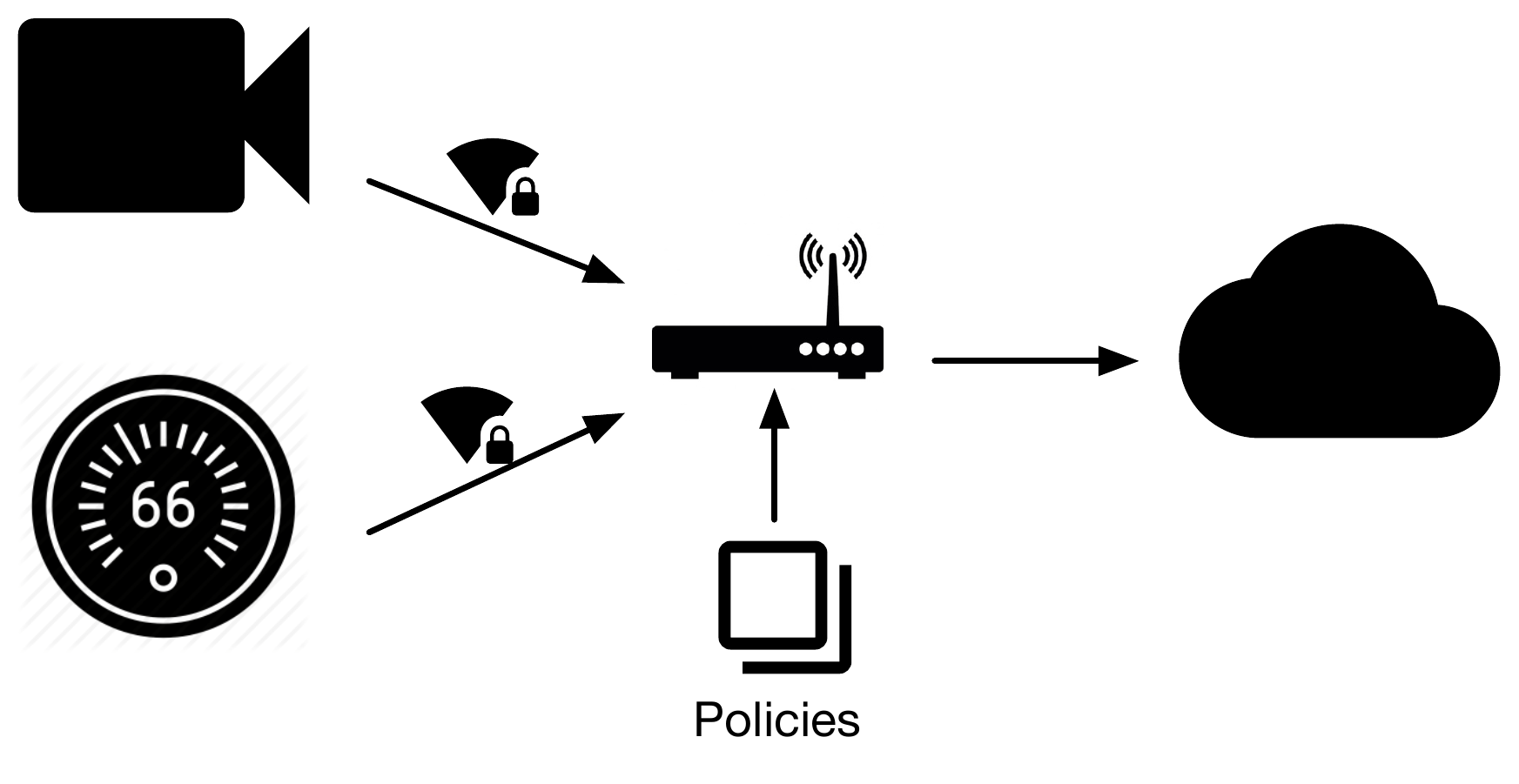}
  \caption{Enforcement at a router}
  \label{fig:overview-router}
\end{subfigure}

\caption{Overview of the policy enforcement architecture. (a) For IoT devices without direct connection to the local network (e.g., certain smart watches and light bulbs), policies are enforced at the network access device (e.g., a smartphone). (b) For IoT devices with direct network connectivity, policies are enforced at the internet gateway (e.g., wireless access point or border router).}
\label{fig:enforcement}
\end{figure}

\begin{figure}
\includegraphics[width=.49\textwidth]{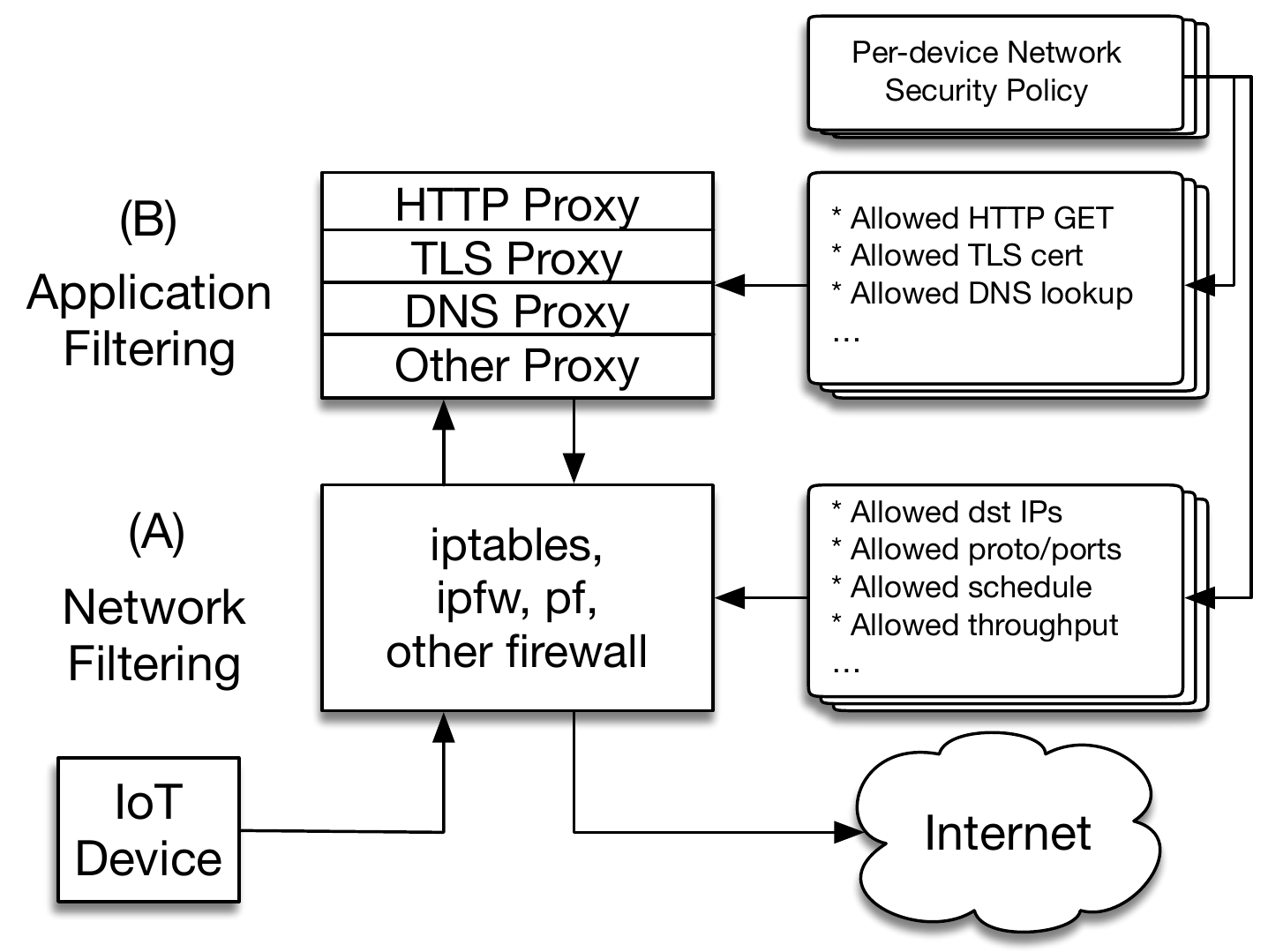}
\caption{Components of the \name policy enforcement architecture.}
\label{fig:overview}
\end{figure}

To be effective at restricting network capabilities, the \name policy enforcement logic must be positioned at a vantage point where all network traffic to and from IoT devices can be inspected. Such vantage points will vary depending on the wireless technology used by the IoT devices, and depending on the topology of the network being instrumented. As shown in Figure~\ref{fig:enforcement}, devices using Zigbee, Bluetooth, and similar short-range network technologies require a hub or smartphone to provide connectivity to cloud services. For these devices, policy enforcement can be built into the hub if supported, or the hub itself may be treated as an Internet-capable IoT device. For devices that support IP connectivity, enforcement can be applied at the wireless access point, WAN gateway, or at a middlebox between the wireless AP and the gateway. We focus filtering on \emph{outgoing} traffic assuming that most home gateways already filter unsolicited inbound traffic (the default behavior when NAT is used). \name could easily be extended to filter inbound traffic as well, in situations where devices are globally reachable (e.g., when using IPv6 with no border firewall).

Once an enforcement point has been selected, a policy describing essential network behavior of the IoT device must be obtained. Enumerating the network behavior that is essential for regular operation of an IoT device can be challenging. For example, should periodic reporting of device usage analytics be considered essential?, How do we distinguish between user-generated network traffic and automated connections? For now, we assume the existence of a policy for each device and in Section~\ref{sec:policies} we discuss strategies for creating, retrieving, and adapting policies to desired behavior.

\name security policies aim to enforce the principle of least privilege. That is, they should describe the minimal set of \emph{allowable} network connections. Policies are therefore whitelists; any connections that do not match at least one entry in the policy will be dropped. Entries may specify connection metadata (e.g., packet sizes, traffic rates, or schedules) or contents (e.g,. IP addresses, ports, protocols, flags). Policies may also specify application layer contents for supported protocols. More details are given in Section~\ref{sec:policies}.

Depending on the type of entries, policies are loaded into one of two filtering modules, as shown in Figure~\ref{fig:overview}. Network layer and metadata entries are loaded into the network filtering module (A). This module converts entries into software firewall rules suitable for use in one of the well-known packet filtering frameworks on Linux, BSD, or other firewall appliances. Application layer entries are loaded into application-layer proxies, each implementing their own enforcement logic. Each proxy inspects payload data from packets, and transparently forwards the request to the destination if the request is compliant with the policy. 

\vspace{3em}
\subsection{Policies}
\label{sec:policies}
\begin{lstlisting}[language=json,firstnumber=1, caption="Example policy for the Netatmo weather station", label=fig:policy]
{"Netatmo Weather Station": {
  "MACAddr": "70:ee:50:13:ab:cd",
  "IPAddr": "172.16.1.2",
  "AllowedDNSQueries": [
    {"type": "A", "query": "netcom.netatmo.net", "resolver": "192.168.1.1"}
  ],
  "AllowedDNSReplies": [
    {"type": "A", "query": "netcom.netatmo.net", "answers": "62.210.92.0/24"}
  ],
  "AllowedConnections": [
    {"family": "IPv4", "dest": "netcom.netatmo.net", "proto": "TCP", "dstport": "25050", "freq": "6/hr"}
  ]
 }
}
\end{lstlisting}

\name policies are machine-readable descriptions of expected network behavior for the IoT device. Policies are whitelists, meaning that any outgoing traffic that is not defined in the policy will be denied. We chose a whitelisting approach instead of blacklisting for two reasons. First, whitelisting, when describing a narrow set of behavior,
provides the strongest security guarantees; it forces an adversary to operate within the confines of rules in the whitelist, as opposed to operating around rules of a blacklist. Second, IoT vendors designing devices should be able to describe how and to what the device needs to connect; while IoT developers may not be security experts, they must be aware of network activity since it is this very activity that gives the device functionality. Because of the whitelisting approach, device policies must ensure the inclusion of rules for all expected connections including periodic API calls, user-triggered network behavior, software/firmware updates, etc. 

Listing~\ref{fig:policy} shows an example policy for the Netatmo weather station. Our analysis of network traces collected for the weather station revealed that the device wakes up every 10 minutes, performs readings of CO$_{2}$, temperature, air quality, and air pressure, and uploads the measurements to Netatmo's cloud servers. To obtain the IP address(es) of Netatmo's servers, the device performs an IPv4 (type \verb|A|) DNS lookup of \verb|netcom.netatmo.net|. The upload takes place over TCP on port 25050 to an IP address returned by the previous DNS lookup. Line 11 in Listing~\ref{fig:policy} concisely captures all the described behavior. It allows outgoing IPv4 TCP connections to port 25050 to any IP address returned by a lookup to \verb|netcom.netatmo.net|, with at most 6 of these connections being initiated per hour (one every 10 minutes). Line 8 restricts the IP addresses that are allowed as answers when performing the DNS lookup, and Line 5 allows lookups of only one domain name via a single resolver. 

Note that the minimal policy in Figure~\ref{fig:policy} appears to be sufficiently restrictive. However, even such a policy could leave room for an attacker to be disruptive. For example, an attacker gaining control of a \name-protected Netatmo weather station could flood the DNS resolver with a large number of A lookups of \verb|netcom.netatmo.net|, or send gigabytes of TCP traffic to any of Netatmo's servers. The policy could be further tightened by specifying additional restrictions such as number of bytes, packets, or number of allowed lookups. Table~\ref{tab:policy} shows additional options that could be defined in the policy. While the table is not meant to be comprehensive, we note that adding a new parameter to the policy only requires a corresponding way to inspect and enforce that parameter at the enforcement point. Linux's \verb|netfilter| framework includes many ways to filter traffic, while new proxies can be written to support new applications and protocols. 

\begin{table}[]
\begin{tabularx}{\linewidth}{lX}
\toprule
Type        & Example Parameters                                                                      \\ \midrule
Metadata    & Schedule, rate, bandwidth, packet size                                                  \\
Contents    & Protocol, IP Address, port number, connection flags/state\\
Application & Types of DNS lookups and responses, TLS certificates, HTTP GET/POST/PUT request \\ \bottomrule
\end{tabularx}
\caption{Example parameters that could be defined in \name policies}
\label{tab:policy}
\end{table}

\subsubsection{Obtaining policies}
\label{sec:obtainingpolicies}

We envision several ways to obtain a policy for a given device. 

\begin{enumerate}[leftmargin=*]
	\item \textbf{Manufacturer.} The device manufacturer can create the policies for devices they ship. We believe manufacturers are in the best position to do so, since they also develop or commission the software for the device. It is thus reasonable to expect the manufacturers to know what functionality the device needs. Policies could be made available through vendor websites (e.g., a QR code on the box pointing to \url{mysmarttoaster.io/securitypolicy}), or distributed along with the software for managing the device. 

	\item \textbf{Third party.} Policies can be written by third parties, either by writing new policies from scratch after understanding the devices behavior, or by modifying manufacturer-provided policies to be more/less restrictive (e.g., by removing/adding rules). The IoT enthusiast community has already enabled integration of vendor-unsupported services and devices (e.g., Homebridge\footnote{\url{https://github.com/nfarina/homebridge}}, Home Assistant\footnote{\url{https://home-assistant.io}}), so they may provide policies for certain devices. Anti-malware and security firms could also provide policies as a service to their customers, creating an additional revenue stream. 

	\item \textbf{Automatic.} If no policies are available from vendors or other experts, it is possible to programatically create policies by observing the network behavior of a device for a given amount of time. The device can be assigned a temporary \emph{allow all} rule, during which all network traffic is recorded. After the monitoring period ends, a policy matching the observed behavior can be created and enforced. 

\end{enumerate}

Once a policy has been retrieved and is being enforced, functionality changes (e.g., through a firmware update) to the IoT device may require updating the previously installed policy. We expect \name will require a mechanism to securely authenticate and verify updated policies. One strategy is to digitally sign policies and verify the validity of the signature against a set of pre-installed trust roots. Alternatively, self-signed certificates along with a trust-on-first-use mechanism (\`{a} la Android~\cite{Barrera2012a}) could be used. We leave policy verification and updates to future work. 

\subsubsection{Human aspects} 
An inevitable consequence of the large number of IoT devices is that non-expert users will become the administrators of dozens of devices. We see \name as a step toward improving transparency of devices, since its machine readable policies can easily be converted to human readable form and displayed on another device (e.g., smartphones or PCs). These policies could allow even non-experts to gain visibility into what their devices are allowed to do. For example, the entries \verb|max-bw-out: 10M/w| and \verb|valid-domains: api.lightbulbs.io| can be converted to: \emph{``This light bulb will not send more than 10 MB of data per week to api.lightbulbs.io''} or simpler \emph{``This light bulb will only connect to api.lightbulbs.io''}.

Another way to offer transparency and visibility is to collect and display statistics at the policy enforcement point. Measurements showing number of times a rule has been matched, or displaying extraneous connections can help identify devices that are misbehaving. 

\section{Proof of concept}
As a proof of concept, we implemented a subset of \name functionality and applied the policy enforcement mechanism to 3 devices: a Netatmo Weather station, a LIFX smart light bulb, and a Fitbit Aria digital scale.

\begin{figure}
\includegraphics[width=.3\textwidth]{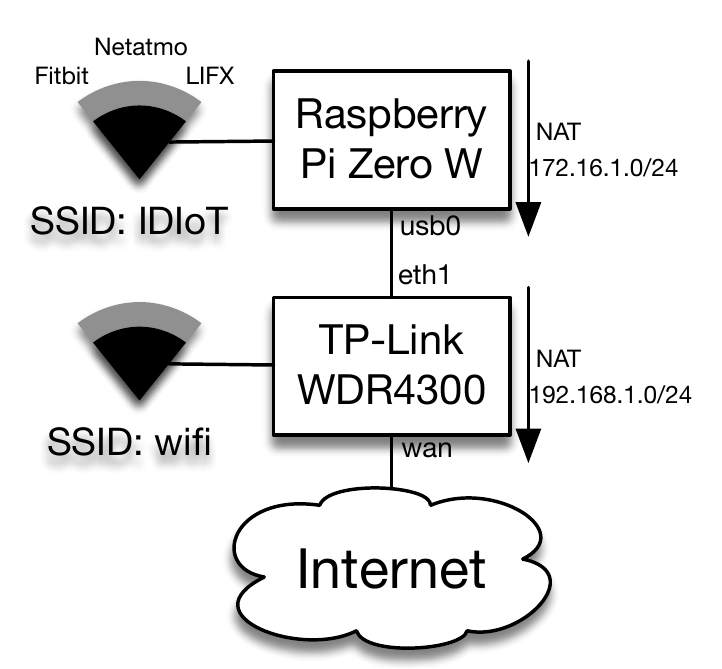}
\caption{Proof of concept \name enforcement. IoT devices connect to the IDIoT wireless network, while other devices (laptops, phones) connect to the wifi network. \name policy enforcement happens at the Raspberry Pi through iptables and dnsmasq.}
\label{fig:poc}
\end{figure}

\subsection{Environment}  

We built a test environment on a Raspberry Pi Zero W running a recent build of Archlinux. The Raspberry Pi acts as a middlebox between IoT devices and the network gateway by advertising a wireless network using \verb|hostapd|\footnote{\url{https://w1.fi/hostapd/}}. A USB-ethernet adapter connects the Raspberry Pi to another router, as shown in Figure~\ref{fig:poc}. This topology allows non-IoT devices to connect to the ``wifi'' wireless network and access the Internet in an unrestricted way. Devices on the \name wireless network are assigned IP addresses through DHCP in the 172.16.1.0/24 network range, while devices on the TP-Link router are on the 192.168.1.0/24 network. Having separate networks and translating between them via NAT offers the additional benefit of effortlessly preventing information leaks about, or rogue connections to the unrestricted network.

Link layer isolation is done by setting \verb|ap_isolate=1| in the \verb|hostapd| configuration. This prevents wireless clients from seeing each other at layer 2. Network layer rules are enforced through \verb|iptables|, Linux's built-in packet filtering framework. DNS filtering is done through \verb|dnsmasq|~\cite{dnsmasq}, a lightweight network infrastructure tool that supports DNS forwarding. 

\subsection{Creating Policies} For each device, we created a policy using the automatic method (see Section~\ref{sec:obtainingpolicies}) and manually inspected the policies for correctness. To automatically create the policies, we developed a tool in \verb|golang| that reads network packet captures (in the form of pcap files) and produces a JSON policy as shown in Listing~\ref{fig:policy}. The tool reads in packets from the capture file, identifying new or ongoing TCP/UDP sessions. The tool also inspects DNS traffic, recording queries and replies. 

For the weather station and the light bulb, we captured network traffic for 12 minutes. Both devices performed some background activity. The digital scale only performs network activity for uploading measurements. Thus, we stepped on the scale twice and recorded network activity. 

Once the packet capture has been read, the tool removes duplicate entries (e.g., recurring DNS queries or TCP connections) and produces a policy object. The object can then be printed to the screen, or written to a file in JSON or other formats. 

\subsection{Enforcement} We wrote a simple tool that takes as input a policy file (see above) and converts entries in the file into \verb|iptables| rules or \verb|dnsmasq| whitelist entries. For example Listing~\ref{fig:policy} is converted to:
\smallskip

\verb|#iptables -t nat -A PREROUTING -i wlan0 \\|

\verb|-s 172.16.1.2 -d 62.210.92.0/24 -p tcp \\|

\verb|--dport 25050 -m limit --limit 6/hour -j ACCEPT|
\smallskip

\verb|#iptables -t nat -A PREROUTING -i wlan0 \\|

\verb|-s 172.16.1.2 -d 192.168.1.1 -p udp \\|

\verb|--dport 53 -j ACCEPT|
\smallskip

Note that our tool combines multiple policy entries (in this case lines 7 and 11) to create a stricter \verb|iptables| rule. Since the connections entry specifies a destination hostname, and there is a corresponding rule specifying allowed IPs for that hostname, the rule can precisely specify allowed sources (-s) and destinations (-d). The second rule allows UDP traffic to destination port 53 as required for DNS lookups. The firewall is configured to drop all traffic that doesn't match at least one rule, and to allow replies to connections that were allowed outbound. 

We configured \verb|dnsmasq| to forward received DNS queries that are allowed by the policy. To do this, the \verb|dnsmasq| configuration file requires the \verb|no-resolv| directive, which instructs the resolver to ignore the system's DNS resolvers, and therefore not do any DNS forwardings unless otherwise specified by a whitelisted entry. These whitelisted entries are extracted from the JSON policies by looking for the ``AllowedLookups'' directive. Allowed lookups are added to the \verb|dnsmasq| configuration file as:
\smallskip

\verb|server=/netcom.netatmo.net/8.8.8.8|
\smallskip

The entry above instructs \verb|dnsmasq| to forward all queries of \verb|netcom.netatmo.net| to Google's public DNS resolver (8.8.8.8). Finally all other lookups (using the wildcard ``\#'') are set to return an address of 127.0.0.1. 
\smallskip

\verb|address=/#/127.0.0.1|
\smallskip

\subsection{Testing} After deploying the enforcement rules, we attempted to use the IoT devices to ensure their functionality was not impaired by our filtering. The Fitbit Aria successfully uploaded weights. The Netatmo weather station was able to perform periodic reporting every 10 minutes, but repeated on-demand readings (triggered by pressing a button on top of the device) made the number of connections exceed the 6/hour threshold. A more permissive value of 10-20 per hour may be more appropriate to allow a small number of on-demand readings. 

The LIFX bulb worked as expected, although with higher latency between commands and responses. The LIFX bulb can be controlled through a smartphone application, which is expected to be on the same local network as the bulb. By having the bulb and smartphone and different networks, commands were sent to LIFX's cloud servers, which were then read by the bulb's long-lived TCP connection to the same servers. While this added latency, it had no effect on functionality.

\section{Discussion}

This section discusses technical challenges in deploying network filtering solutions to secure IoT devices, and outlines non-networking issues in the IoT security domain. 

\subsection{Technical Challenges for \name and \name-like Solutions}

\subsubsection{Device-to-device connectivity} Certain IoT devices require discovery and connectivity to other devices on the network. In particular, devices that don't rely on cloud services may operate by discovering nearby devices and interact with them directly. While disabling access point isolation may enable certain use cases, it also opens up the all devices on the wireless network to attacks. There may be opportunities for ``selective AP isolation'', where devices can be allowed to communicate with authorized specific devices on the same network. 

\subsubsection{Device Identification.} Current IP networks identify devices based on layer 2 identifiers (MAC addresses) and IP addresses. When creating or loading a policy that applies to a given device, it is still possible for a compromised device to modify its behavior, and simultaneously modify its identifiers. Miettinen et al.~\cite{miettinen_iot_2017} show that fingerprinting device types can be done with high accuracy, but identifying distinct firmware builds or hardware variants of the same device is more challenging. Being unable to identify a device correctly could allow a device to spoof the behavior of a different device with a less restrictive policy. While better fingerprinting techniques are developed, an alternative solution to this problem is remote attestation, but this requires a trusted hardware module.

\subsubsection{Complex IoT devices} Throughout the paper, we've described how our proposed policy enforcement framework can be effective when devices have a small set of predictable functionality. Given the rapid pace of innovation in IoT, it is reasonable to expect IoT devices to grow in complexity. As devices gain features that allow customization or extensibility, our ability to profile and restrict them drops. This is already the case for IoT-ish devices like the Xbox or the AppleTV (see Table~\ref{tab:summary}).  These multimedia boxes allow the installation of applications, blurring the line between single-purpose functionality and general purpose computers. Because each new application may require connecting to a variety of cloud services, enumerating all possible servers and protocols may become infeasible. Personal desktop firewalls experienced usability challenges for this very reason over a decade ago; repeated prompts to allow network connectivity for each new application were often dealt with allowing all outbound connections~\cite{herzog_usability_2007}.

\subsubsection{WAN-enabled IoT devices} As wireless technology costs decrease, manufacturers may start shipping products with built-in WAN connectivity. Direct WAN connectivity increases usability by removing the need for complex network attachment procedures, and also gives vendors direct access to the device for diagnostics and updates. The downside of direct WAN connectivity is of course the consumer's inability to control the communication channel. IoT devices with such capabilities already exist; for example, the Amazon Kindle can download books and updates over its built-in 3G connection. Another emerging technology is LoRaWAN~\cite{lorawan}, a low-power wireless protocol which allows devices to effortlessly join city-wide networks. 

\subsection{Beyond \name}

We've described \name as a possible solution to the IoT security problem. We've shown that \name can be effective in restricting what compromised devices do on the network, which protects both internal and external hosts from attack. However, network-based attacks are not the only threat to IoT devices and users. 

\subsubsection{Data Privacy} Data produced by IoT devices can be stored on cloud servers for facilitating interaction with other services or for displaying to users through web interfaces. Network filtering techniques can only ensure that the data is sent to the expected endpoints, but cannot enforce what the data is used for by the cloud provider. While promising approaches for preserving data privacy exist (e.g., differential privacy~\cite{dwork2008differential} and homomorphic encryption~\cite{naehrig2011can}, these approaches have yet to see broad adoption. 

\subsubsection{Device/Service longevity} Certain classes of IoT devices have long expected lifespans. For example, LIFX advertises that their bulbs should last around 22 years if used 3 hours per day. It may be unreasonable to expect companies to provide security updates to devices for such long periods, but long device lifespans pose a number of (sometimes non-technical) challenges. What happens with user data and devices if the company is acquired or files for bankruptcy? What happens if domain names for API endpoints or IP address space move to a new owner? IoT vendors must consider these cases and design their infrastructure accordingly. If the decision is to plan the obsolescence of devices (c.f. Revolv~\cite{revolv}), users should be informed. 

\section{Conclusion}

This paper discusses the challenges in securing billions of consumer IoT devices. We argue that security solutions requiring vendor involvement, such as modifications to hardware and software are unlikely to be successful. We propose \name, a network-based isolation and filtering system for IoT devices. Our proof-of-concept showed that Mirai-style attacks can be prevented without any modifications to devices. The design of \name demonstrates that there are simple problems buried within a complicated paradigm, and these can be solved effectively without over-engineered solutions. We argue that additional straightforward solutions are needed to help secure IoT devices in years to come. 

\section*{Acknowledgements}
We thank Andreas Wespi, Anton Beitler, and members of the security group at IBM Research Zurich for their insightful discussions. We also thank Paul van Oorschot for feedback on early versions of this paper.

\bibliographystyle{ACM-Reference-Format}
\bibliography{bibliography} 

\end{document}